\documentclass[a4paper,11pt]{article}
\usepackage{pos}

\title{The Machine Learning to reconstruct GRB lightcurves}

\author*[1,2,3,4]{M. G. Dainotti}
\author[5,6]{B. De Simone}
\author[7,8]{A. Narendra}
\author[8,9]{A. Pollo}

\affiliation[1]{Division of Science,National Astronomical Observatory of Japan, 2 Chome-21-1 Osawa, Mitaka, Tokyo, 181-8588, Japan}
\affiliation[2]{The Graduate University for Advanced Studies, SOKENDAI, Shonankokusaimura, Hayama, Miura District, Kanagawa, 240-0115, Japan}
\affiliation[3]{Space Science Institutes , 4765 Walnut St Ste B, Boulder, 80301, CO, USA}
\affiliation[4]{Nevada Center for Astrophysics, University of Nevada,89154, 4505 Maryland Parkway, Las Vegas, 80301, NV, USA}
\affiliation[5]{Dipartimento di Fisica “E.R. Caianiello”, Università di Salerno, Via Giovanni Paolo II, 132, Fisciano, Salerno, 84084, Italy}
\affiliation[6]{INFN Gruppo Collegato di Salerno- Sezione di Napoli. c/o Dipartimento di Fisica ”E.R. Caianiello”, Università di Salerno, Via Giovanni Paolo II, 132, Fisciano, Salerno, 84084, Italy}
\affiliation[7]{Doctoral School of Exact and Natural Sciences, Jagiellonian University, Krakow, Poland}
\affiliation[8]{Astronomical Observatory of Jagiellonian University, Krakow, Poland}
\affiliation[9]{National Center for Nuclear Physics (NCB), Warsaw}

\emailAdd{maria.dainotti@nao.ac.jp}

\abstract{The current knowledge in cosmology deals with open problems whose solutions are still under investigation. The main issue is the so-called Hubble constant ($H_0$) tension, namely, the $4-6 \sigma$ discrepancy between the local value of $H_0$ obtained with Cepheids+Supernovae Ia (SNe Ia) and the cosmological one estimated from the observations of the Cosmic Microwave Background (CMB). For the investigation of this problem, probes that span all over the redshift $z$ ranges are needed. Cepheids are local objects, SNe Ia reached up to $z=2.9$, and CMB is observed at $z=1100$. In this context, the use of probes at intermediate redshift $z>3$ is auspicious for casting more light on modern cosmology. The Gamma-ray Bursts (GRBs) are particularly suitable for this task, given their observability up to $z=9.4$. The use of GRBs as standardizable candles requires the use of tight and reliable astrophysical correlations and the presence of gaps in the GRB time-domain data represents an obstacle in this sense. In this work, we propose to improve the precision of the lightcurve (LC) parameters through a reconstruction process performed with the functional forms of GRB LCs and the Gaussian Processes (GP). The filling of gaps in the GRB LCs through these processes shows an improvement up to $41.5\%$ on the precision of the LC parameters fitting, which lead to a reduced scatter in the astrophysical correlations and, thus, in the estimation of cosmological parameters.}

\FullConference{Frontier Research in Astrophysics – IV (FRAPWS2024)\\
 9-14 September 2024\\
Mondello, Palermo, Italy\\}

\begin{document}
\maketitle

\section{Introduction}
The most studied cosmological framework is the so-called flat $\Lambda$CDM model, which lies its basis on the existence of Dark Energy (parametrized with $\Lambda$) responsible for the expansion of the universe, the Cold Dark Matter component (which is non-relativistic), and the absence of a geometrical curvature. This model has proven to be one of the best paradigms for describing the cosmos, given its capability of predicting an accelerated expansion phase as proven in \cite{27,22} thanks to the crucial contribution of SNe Ia. These are key observational pillars, together with the CMB anisotropy observations. The $\Lambda$CDM is then the most popular cosmological model among scientists. Nevertheless, many open problems affect the $\Lambda$CDM. A major challenge in cosmology today is the Hubble tension - a discrepancy in measurements of the Hubble constant ($H_0$). This parameter quantifies the universe's current expansion rate. Local measurements using SNe Ia and Cepheids yield a higher 
$H_0$ value ($H_0=73.04 \pm 1.04$ \cite{28}) than predictions derived from the early universe, based on CMB observations by the Planck satellite ($H_0=67.4 \pm 0.5$ \cite{24}).
The presence of such a big open problem in modern cosmology calls out the need for further constraints to be put on the $H_0$ estimation. In particular, it is important to cover the redshift ($z$) ranges in the Hubble diagram that lie between the SNe Ia upper limit ($z=2.9$ \cite{23}) and the value of the Last Scattering Surface ($z=1100$) from which CMB is emitted. To this end, the contribution of GRBs is crucial. GRBs are highly energetic and panchromatic transients observed isotropically in the universe and have cosmological origins (being observed up to $z=9.4$ \cite{4}). GRB LCs have two main emission phases: the prompt, observed in high-energy frequencies ($\gamma,\,X$, optical) and due to internal shocks mechanism, and the afterglow (usually observed in $X,$ optical, and occasionally radio) which is generated from the interaction of the shock shells with the interstellar medium. GRBs are traditionally divided into two categories according to their emission duration: the Long GRBs (with emission time $>2\,sec$), thought to arise from the collapse of massive stars and the Short GRBs (with a duration typically $<2\,sec$) that are created when two compact objects merge, like two neutron stars and a neutron star with a black hole. Differently from SNe Ia, the GRB emission energies vary widely ($10^{49}-10^{53}\,ergs$), thus to use them as standardizable candles it is fundamental to consider reliable astrophysical correlations among their luminosity and other parameters that do not depend on the luminosity itself. To this end, a particular class of GRBs, namely the GRBs with plateau emission, serves the purpose. The plateau is a flat portion present in $42\%$ of the GRB LC which is related to the fallback accretion mechanism onto a black hole or the spinning down of a newly born magnetar \cite{29,26}. The plateau phase is of particular importance given that, morphologically, it is more regular than the prompt phase in the LCs. Furthermore, the plateau is the protagonist in reliable astrophysical correlations aimed at the standardizability of GRBs. Among these relations, the \emph{fundamental plane} or \emph{Dainotti 3D relation} serves as a reliable cosmological tool: this correlation connects the GRBs peak prompt luminosity $L_{peak}$, the end-of-plateau GRB luminosity and rest-frame time, $L_{a}$ and $T^{*}_{a}$, respectively \cite{5, 15, 16, 14, 10, 6, 7, 30, 12}. The fundamental plane relation can provide interesting support in the cosmological analysis \cite{25, 3, 9, 1, 11, 17, 2}, as well as quasars as high-$z$ probes \cite{20}.
To have more reliable cosmological models, we need to have the tightest possible constraints and this can be obtained when we have the smallest uncertainties on the plateau parameters in relation to the Dainotti 3D correlation.

For the reasons stated above, it is clear that GRBs, together with other high-$z$ probes, represent the developing future of modern cosmology. Nevertheless, the issues with these probes are still challenging the scientific community. One of the main problems concerns the \emph{lack of data}: the $X$-ray satellite observations may incur in the so-called orbital gaps, i.e. phases of the GRB observation where data can't be collected due to the satellite orbital gap. Furthermore, concerning the optical wavelengths, not all the observatories on Earth that are triggered from the satellites have the possibility to observe the GRB location at the same time, thus, LCs may present fragmented contributions at different time epochs \cite{8}.

In this sense, leveraging an approach that allows \emph{filling the gaps} increases drastically the precision of the fitting parameters for a GRB LC. An improvement in GRB parameters precision implies a reduction in cosmological parameters uncertainties when these transients are used as standardizable candles.

The main idea of the present work is to use \emph{machine learning approaches} to fill the gaps present in GRB LCs. The reconstruction approach has been applied with success in many fields of modern astrophysics, such as the study of Cepheid LCs \cite{21} or planetary eclipse mapping \cite{19}.
The present contribution is structured in the following way. In Section \ref{LCR} we describe the LC reconstruction process applied on $X$-ray GRBs with plateau emission. In Section \ref{conclusions}, we draw our conclusions.

\section{GRB lightcurve reconstruction}\label{LCR}
The problem of GRB LC reconstruction has been investigated in \cite{13}, where the authors address the challenge of gaps in GRB LCs. Starting from a sample of 455 GRBs taken from \cite{31} that have $X$-ray plateaus observed by Swift \cite{18}, a stochastic approach is followed to reconstruct the missing data points in the LC gaps. 
After the data points are reconstructed and the LCs filled, a re-fitting of the LCs with two theoretical models is performed to compare the precision of the LC parameters before and after the reconstruction process.
Two main methods are used to infer the missing data points: \emph{reconstruction with functional forms} and \emph{reconstruction with Gaussian Processes (GP)}.\\

\subsection{Reconstruction with the functional form}
The 218 good GRBs are fitted with the Willingale model \cite{32}:

\begin{equation}
f(t)=\begin{cases}
			F_i exp\left(\alpha_i \left(1-\frac{t}{T_i} \right) \right) exp\left(\frac{-t_i}{t}\right), & \text{for $t<T_i$}\\
             F_i \left( \frac{t}{T_i} \right)^\alpha_i exp\left(\frac{-t_i}{t} \right), & \text{for $t \geq T_i$}
		 \end{cases},
         \label{eq:Willingale}
\end{equation}

where $F_i$ and $T_i$ are the flux and time either at the end of the plateau or the prompt phase of GRB LC, $\alpha_i$ is the temporal index after $T_i$, and $t_i$ is the initial time rise.
Another model taken into account by the authors is the simple broken power-law, defined as follows:

\begin{equation}
f(t)=\begin{cases}
			F_i \left(\frac{t}{T_i} \right)^\alpha_1, & \text{for $t<T_i$}\\
             F_i \left(\frac{t}{T_i} \right)^\alpha_2, & \text{for $t \geq T_i$}
		 \end{cases}
         \label{eq:BPL}
\end{equation}

where $T_i$ and $F_i$ are the flux at the breaking point (which is taken as the plateau in this context), $\alpha_1$ is the slope before the break, and $\alpha_2$ is the one after the break. 
For each LC, the residuals with respect to the logarithm of the models are defined:

\begin{equation}
    \log_{10}F_{res}=\log_{10}F^{obs}_{t}-\log_{10}f(t),
    \label{eq:residuals}
\end{equation}

considering the $\log_{10}f(t)$ as the fitting model and $\log_{10}F^{obs}_{t}$ as the observed $\log_{10}$ of flux at time $t$. A test of the best-fit distribution for the model residuals is performed, confirming the Gaussian nature for them. Before reconstructing the GRB LCs, a segregation of the 455 GRBs into quality categories is necessary to understand which ones can be reconstructed. The \emph{good GRBs} are eligible for reconstruction: the ones that are well approximated by fitting models and do not have any flare, bump or double break in the plateau or afterglow phase. In this way, 218 GRBs (48\% of the total) are chosen for the present analysis.

The reconstruction of the LCs is performed through the following Equation:

\begin{equation}
    \log_{10}F^{rec}_{t}=\log_{10}f(t)+(1+m)\cdot\mathcal{R},
    \label{eq:recon}
\end{equation}

where $m$ is the noise level and $\mathcal{R}$ is the random variate sampled from a Gaussian distribution. In this sense, white noise is added to take into account the realistic fluctuations in an observed LC. The noise level can be toggled arbitrarily and here it is considered as $10\%$ or $20\%$. We here stress that the reconstruction is performed at the time epoch between the beginning of the plateau emission and the end of the afterglow, namely, the end of the LC.

\subsection{Reconstruction with the Gaussian Processes}
\begin{figure}[h]
    \centering
    \includegraphics[scale=0.7]{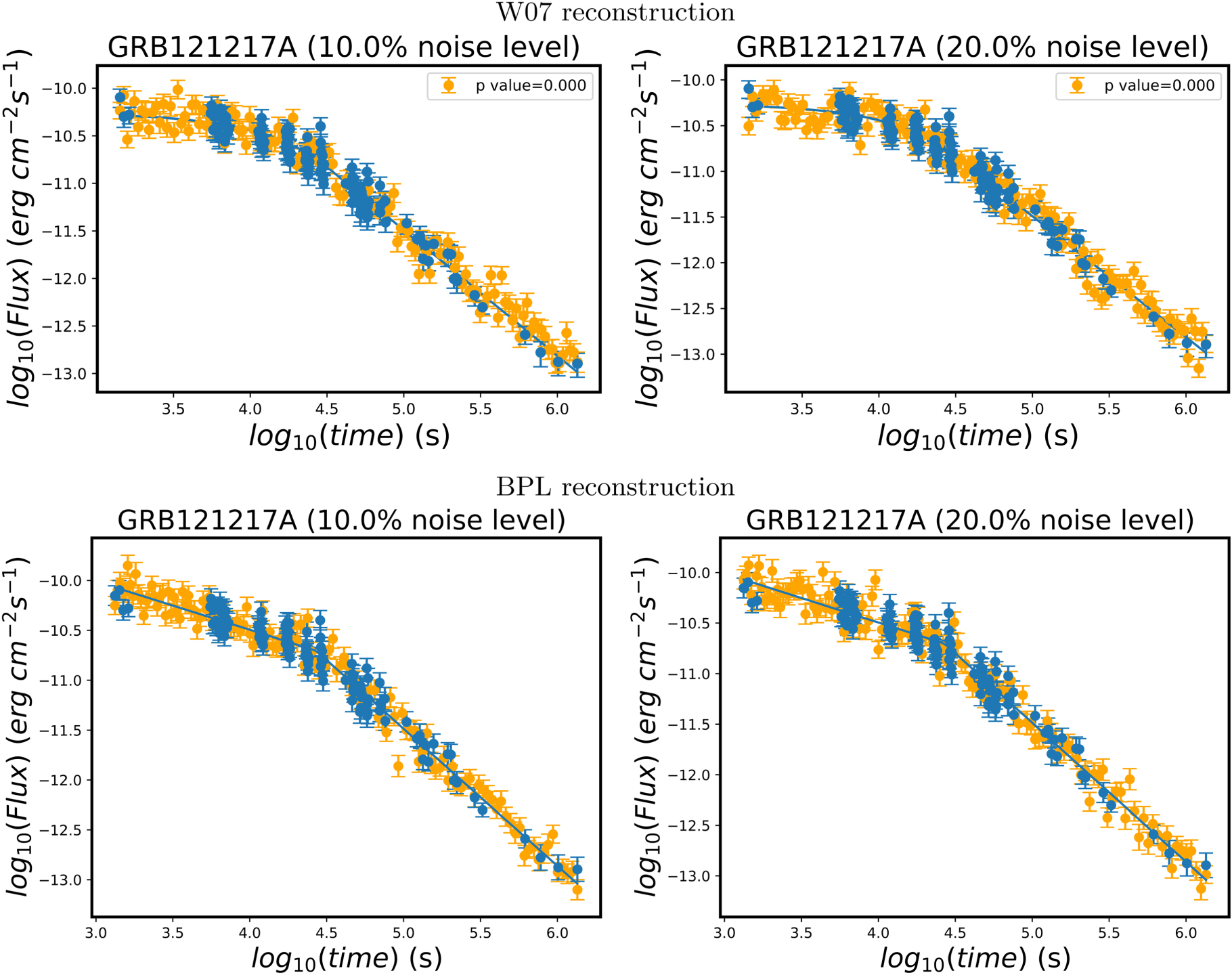}
    \caption{An example of LC reconstruction applied to GRB 121217A. The reconstruction is performed with two models: Willingale (abbreviated with W07) and broken power-law (abbreviated with BPL). Furthermore, the two noise levels are considered ($10\%$ and $20\%$). \textbf{Upper left panel:} W07 model reconstruction with $10\%$ noise level. \textbf{Upper right panel:} W07 model reconstruction with $20\%$ noise level. \textbf{Lower left panel:} BPL model reconstruction with $10\%$ noise level. \textbf{Lower right panel:} BPL model reconstruction with $20\%$ noise level.
    The Figure is reproduced from \cite{13}.}
    \label{fig:reconstruction}
\end{figure}

The Gaussian Processes (GPs) are a class of supervised machine learning methods aimed to perform the regression of data. Starting from the prior knowledge, namely any trend that is found in the existing data used for training, the GP creates a posterior distribution using the Bayesian inference as a principle. The so-found posterior is a likely outcome of the process that is in agreement with both the data and the prior. Every prediction obtained through the GPs has a confidence interval, and the GPs take into account the likelihood of each prediction obtained in the process.
In the GPs, the kernel functions are crucial: these describe how similar are two data points in the input and it is based on the assumption that two similar data points in the input space should lead to similar outputs. In this work, the kernel is assumed as a Radial Basis Function (RBF), which depends only on the difference of any two input data points $x$ and $x'$, thus being a function of $|x-x'|$. Furthermore, the 95\% confidence interval is assumed for the GPs analysis. An example of LC reconstruction is plotted in Figure \ref{fig:reconstruction}.

\section{The results of the reconstruction}
After the reconstruction and the re-fitting with the models in Equations \ref{eq:Willingale}, \ref{eq:BPL} are employed, the following observations can be made. The decrease in percentage is a measurement of how the fitting has improved after the reconstruction process. It is defined as

\begin{equation}
    \Delta_{\% X} = \frac{\epsilon f^{a}_{X}-\epsilon f^{b}_{X}}{\epsilon f^{b}_{X}}*100,
\end{equation}

where $\epsilon f^{b}_{x}=|\Delta_x/x|_{before}$ is the error fraction of the $x$ LC parameter before the reconstruction and $f^{a}_{x}=|\Delta_x/x|_{after}$ is the same quantity but after the reconstruction. The $\Delta_\%$ is computed for each GRB and then an average is estimated among all the GRBs.

We here report the results on the $\Delta_\%$ after the reconstruction with the functional form of the fitting functions. In the case of $m=10\%$ noise level, considering the average among all GRBs, we obtain a 33\% decrease for $\Delta_\% \log_{10}T^{*}_{a}$, a 31\% decrease for $\Delta_\% \log_{10}F_{a}$, a 15\% decrease for $\Delta_\% \alpha_1$ parameter, and a 44\% decrease for $\Delta_\% \alpha_2$. For what it concerns the $m=20\%$ noise level, we obtain a 30\% decrease for $\Delta_\% \log_{10}T^{*}_{a}$, a 27\% decrease for $\Delta_\% \log_{10}F_{a}$, a 2\% decrease for $\Delta_\% \alpha_1$, and a 41\% decrease for $\Delta_\% \alpha_2$.

Concerning the GP results, the summary follows. For $\Delta_\% \log_{10}T^{*}_{a}$, the decrease is of 25\%. For $\Delta_\% \log_{10}F_{a}$ and $\Delta_\% \alpha$, the decreases are 28\% and 42\%, respectively. Furthermore, we observe a decrease in $\Delta_\% \log_{10}T^{*}_{a}$ of 15\%, in $\Delta_\% \log_{10}F_{a}$ of 12\%, in $\Delta_\% \log_{10} \alpha_1$ of 25\%, and in $\Delta_\% \log_{10} \alpha_2$ of 36\%.

In both the functional and GP reconstruction approaches, the results between the Willingale and the broken power-law are very similar. Furthermore, the GP technique exhibited similar trends as the cases with functional forms, reinforcing the effectiveness of the stochastic reconstruction approach.

\section{Conclusions}\label{conclusions}
The methodology proposed in this work aims to solve in a relatively simple way the problem of missing data in the GRB LCs. Through the LC reconstruction and the imputing of data, an improvement in GRB LC fitting parameters can be easily obtained. Here the Willingale and the broken power-law models are considered as the starting points for the analysis but it must be kept in mind that any LC empirical model can be adopted with this approach. We obtain that - on average - for the Willingale model the uncertainties on the fitting parameters are reduced by 37\% for all the parameters with noise at $10\%$ and by 34\% for the $20\%$ noise case, while for the broken power-law the average reduction on the uncertainties of the fitting parameters is of 31\% for the $10\%$ noise level and 25\% for the $20\%$ noise case. Discussing the results on the GPs, the average decrease is 31\% for the Willingale model and 22\% for the broken power-law. This stochastic reconstruction method enhances the accuracy of GRB LC analyses by reducing uncertainties in critical LC parameters. These improvements are vital for utilizing GRBs as standard candles in cosmology, investigating theoretical models, and inferring GRB redshifts through future machine-learning analyses.



\begin{thebibliography}{99}
\bibitem{1} G. Bargiacchi et al. “Gamma-ray bursts, quasars, baryonic acoustic os-
cillations, and supernovae Ia: new statistical insights and cosmological
constraints”. In: Monthly Notices of the Royal Astronomical Society 521.3
(May 2023), pp. 3909–3924. doi: 10.1093/mnras/stad763. arXiv: 2303.
07076 {astro-ph.CO}

\bibitem{2} Giada Bargiacchi, Maria Giovanna Dainotti, and Salvatore Capozziello.
“High-redshift cosmology by Gamma-Ray Bursts: An overview”. In: New
Astronomy Reviews 100, 101712 (June 2025), p. 101712. doi: 10.1016/
j.newar.2024.101712. arXiv: 2408.10707 {astro-ph.CO}

\bibitem{3} Shulei Cao, Maria Dainotti, and Bharat Ratra. “Gamma-ray burst data
strongly favour the three-parameter fundamental plane (Dainotti) cor-
relation over the two-parameter one”. In: Monthly Notices of the Royal
Astronomical Society 516.1 (Oct. 2022), pp. 1386–1405. doi: 10.1093/
mnras/stac2170. arXiv: 2204.08710 {astro-ph.CO}

\bibitem{4} A. Cucchiara et al. “A Photometric Redshift of z ˜9.4 for GRB 090429B”.
In: The Astrophysical Journal 736.1, 7 (July 2011), p. 7. doi: 10.1088/
0004-637X/736/1/7. arXiv: 1105.4915 {astro-ph.CO}

\bibitem{5} M. G. Dainotti, V. F. Cardone, and S. Capozziello. “A time-luminosity
correlation for $\gamma$-ray bursts in the X-rays”. In: Monthly Notices of the
Royal Astronomical Society 391.1 (Nov. 2008), pp. L79–L83. doi: 10 .
1111/j.1745-3933.2008.00560.x. arXiv: 0809.1389 {astro-ph}

\bibitem{6} M. G. Dainotti et al. “A Fundamental Plane for Long Gamma-Ray Bursts
with X-Ray Plateaus”. In: The Astrophysical Journal Letters 825.2, L20
(July 2016), p. L20. doi: 10.3847/2041-8205/825/2/L20. arXiv: 1604.
06840 {astro-ph.HE}

\bibitem{7} M. G. Dainotti et al. “A Study of the Gamma-Ray Burst Fundamental
Plane”. In: The Astrophysical Journal 848.2, 88 (Oct. 2017), p. 88. doi:
10.3847/1538-4357/aa8a6b. arXiv: 1704.04908 {astro-ph.HE}

\bibitem{8} M. G. Dainotti et al. “An optical gamma-ray burst catalogue with mea-
sured redshift - I. Data release of 535 gamma-ray bursts and colour evolu-
tion”. In: Monthly Notices of the Royal Astronomical Society 533.4 (Oct.
2024), pp. 4023–4043. doi: 10.1093/mnras/stae1484. arXiv: 2405.02263
{astro-ph.HE}

\bibitem{9} M. G. Dainotti et al. “Optical and X-ray GRB Fundamental Planes as
cosmological distance indicators”. In: Monthly Notices of the Royal Astro-
nomical Society 514.2 (Aug. 2022), pp. 1828–1856. doi: 10.1093/mnras/
stac1141. arXiv: 2203.15538 {astro-ph.CO}

\bibitem{10} M. G. Dainotti et al. “Selection Effects in Gamma-Ray Burst Correlations:
Consequences on the Ratio between Gamma-Ray Burst and Star Forma-
tion Rates”. In: The Astrophysical Journal 800.1, 31 (Feb. 2015), p. 31.
doi: 10.1088/0004-637X/800/1/31. arXiv: 1412.3969 {astro-ph.HE}

\bibitem{11} M. G. Dainotti et al. “The gamma-ray bursts fundamental plane correla-
tion as a cosmological tool”. In: Monthly Notices of the Royal Astronomical
Society 518.2 (Jan. 2023), pp. 2201–2240. doi: 10.1093/mnras/stac2752.
arXiv: 2209.08675 {astro-ph.HE}

\bibitem{12} M. G. Dainotti et al. “The Optical Two- and Three-dimensional Funda-
mental Plane Correlations for Nearly 180 Gamma-Ray Burst Afterglows
with Swift/UVOT, RATIR, and the Subaru Telescope”. In: The Astro-
physical Journals 261.2, 25 (Aug. 2022), p. 25. doi: 10 . 3847 / 1538 -
4365/ac7c64. arXiv: 2203.12908 {astro-ph.HE}

\bibitem{13} Maria G. Dainotti et al. “A Stochastic Approach to Reconstruct Gamma-
Ray-burst Light Curves”. In: The Astrophysical Journals 267.2, 42 (Aug.
2023), p. 42. doi: 10 . 3847 / 1538 - 4365 / acdd07. arXiv: 2305 . 12126
{astro-ph.HE}

\bibitem{14} Maria Giovanna Dainotti et al. “Determination of the Intrinsic Luminosity
Time Correlation in the X-Ray Afterglows of Gamma-Ray Bursts”. In:
The Astrophysical Journal 774.2, 157 (Sept. 2013), p. 157. doi: 10.1088/
0004-637X/774/2/157. arXiv: 1307.7297 {astro-ph.HE}

\bibitem{15} Maria Giovanna Dainotti et al. “Discovery of a Tight Correlation for
Gamma-ray Burst Afterglows with “Canonical” Light Curves”. In: The
Astrophysical Journal Letters 722.2 (Oct. 2010), pp. L215–L219. doi: 10.
1088/2041-8205/722/2/L215. arXiv: 1009.1663 {astro-ph.HE}

\bibitem{16} Maria Giovanna Dainotti et al. “Study of Possible Systematics in the L*X -
T*a Correlation of Gamma-ray Bursts”. In: The Astrophysical Journal
730.2, 135 (Apr. 2011), p. 135. doi: 10.1088/0004- 637X/730/2/135.
arXiv: 1101.1676 {astro-ph.HE}

\bibitem{17} Arianna Favale et al. “Towards a new model-independent calibration of
Gamma-Ray Bursts”. In: Journal of High Energy Astrophysics 44 (Nov.
2024), pp. 323–339. doi: 10.1016/j.jheap.2024.10.010. arXiv: 2402.
13115 {astro-ph.CO}

\bibitem{18} N. Gehrels et al. “The Swift Gamma-Ray Burst Mission”. In: The Astro-
physical Journal 611.2 (Aug. 2004), pp. 1005–1020. doi: 10.1086/422091.
arXiv: astro-ph/0405233 {astro-ph}

\bibitem{19} Huber, K. F. et al. “Planetary eclipse mapping of CoRoT-2a - Evolution,
differential rotation, and spot migration”. In: Astronomy \& Astrophysics
514 (2010), A39. doi: 10 . 1051 / 0004 - 6361 / 200913914. url: https :
//doi.org/10.1051/0004-6361/200913914.

\bibitem{20} Aleksander Lukasz Lenart et al. “A Bias-free Cosmological Analysis with
Quasars Alleviating H 0 Tension”. In: The Astrophysical Journal Supple-
ment 264.2, 46 (Feb. 2023), p. 46. doi: 10 . 3847 / 1538 - 4365 / aca404.
arXiv: 2211.10785 {astro-ph.CO}

\bibitem{21} Chow-Choong Ngeow et al. “Reconstructing a cepheid light curve with
Fourier techniques. I. The Fourier expansion and interrelations”. In: The
Astrophysical Journal 586.2 (2003), p. 959.

\bibitem{22} S. Perlmutter et al. “Measurements of $\Omega$ and $\Lambda$ from 42 High-Redshift
Supernovae”. In: The Astrophysical Journal 517.2 (June 1999), pp. 565–
586. doi: 10.1086/307221. arXiv: astro-ph/9812133 {astro-ph}

\bibitem{23} J. D. R. Pierel et al. “Discovery of an Apparent Red, High-velocity Type
Ia Supernova at z = 2.9 with JWST”. In: The Astrophysical Journall
971.2, L32 (Aug. 2024), p. L32. doi: 10.3847/2041-8213/ad6908. arXiv:
2406.05089 {astro-ph.GA}

\bibitem{24} Planck Collaboration et al. “Planck 2018 results. VI. Cosmological pa-
rameters”. In: Astronomy \& Astrophysics 641, A6 (Sept. 2020), A6. doi:
10.1051/0004-6361/201833910. arXiv: 1807.06209 {astro-ph.CO}

\bibitem{25} Purva Raut and Apurva Dani. “Correlation Between Number of Hidden
Layers and Accuracy of Artificial Neural Network”. In: Advanced Comput-
ing Technologies and Applications. Ed. by Hari Vasudevan et al. Singapore:
Springer Singapore, 2020, pp. 513–521. isbn: 978-981-15-3242-9.

\bibitem{26} N. Rea et al. “Constraining the GRB-Magnetar Model by Means of the
Galactic Pulsar Population”. In: The Astrophysical Journal 813.2, 92
(Nov. 2015), p. 92. doi: 10.1088/0004- 637X/813/2/92. arXiv: 1510.
01430 {astro-ph.HE}

\bibitem{27} Adam G. Riess et al. In: The Astronomical Journal 116.3 (Sept. 1998),
pp. 1009–1038. doi: 10.1086/300499. arXiv: astro-ph/9805201 {astro-ph}

\bibitem{28} Adam G. Riess et al. “A Comprehensive Measurement of the Local Value of the Hubble Constant with 1 km s-1 Mpc-1 Uncertainty from the Hubble Space Telescope and the SH0ES Team”. In: The Astrophysical Journal
934.1, L7 (July 2022), p. L7. doi: 10.3847/2041- 8213/ac5c5b. arXiv:
2112.04510 {astro-ph.CO}

\bibitem{29} A. Rowlinson et al. “Constraining properties of GRB magnetar central
engines using the observed plateau luminosity and duration correlation”.
In: Monthly Notices of the Royal Astronomical Society 443.2 (Sept. 2014),
pp. 1779–1787. doi: 10.1093/mnras/stu1277. arXiv: 1407.1053 {astro-ph.HE}

\bibitem{30} G. P. Srinivasaragavan et al. “On the Investigation of the Closure Re-
lations for Gamma-Ray Bursts Observed by Swift in the Post-plateau
Phase and the GRB Fundamental Plane”. In: The Astrophysical Journal
903.1, 18 (Nov. 2020), p. 18. doi: 10.3847/1538- 4357/abb702. arXiv:
2009.06740 {astro-ph.HE}

\bibitem{31} G. P. Srinivasaragavan et al. “On the Investigation of the Closure Re-
lations for Gamma-Ray Bursts Observed by Swift in the Post-plateau
Phase and the GRB Fundamental Plane”. In: The Astrophysical Jour-
nal 903.1 (Oct. 2020), p. 18. doi: 10 . 3847 / 1538 - 4357 / abb702. url:
https://dx.doi.org/10.3847/1538-4357/abb702.

\bibitem{32} R. Willingale et al. “Testing the Standard Fireball Model of Gamma-
Ray Bursts Using Late X-Ray Afterglows Measured by Swift”. In: The
Astrophysical Journal 662.2 (June 2007), pp. 1093–1110. doi: 10.1086/
517989. arXiv: astro-ph/0612031 {astro-ph}
\end{thebibliography}
\end{document}